# TacShade: A New 3D-printed Soft Optical Tactile Sensor Based on Light, Shadow and Greyscale for Shape Reconstruction

Zhenyu Lu*, Jialong Yang*, Haoran Li, Yifan Li, Weiyong Si,
Nathan Lepora, *Member, IEEE* and Chenguang Yang †, *Fellow, IEEE*

*Abstract—* In this paper, we present the TacShade: a newly designed 3D-printed soft optical tactile sensor. The sensor is developed for shape reconstruction under the inspiration of sketch drawing that uses the density of sketch lines to draw light and shadow, resulting in the creation of a 3D-view effect. TacShade, building upon the strengths of the TacTip, a single-camera tactile sensor of large in-depth deformation and being sensitive to edge and surface following, improves the structure in that the markers are distributed within the gap of papillae pins. Variations in light, dark and grey effects can be generated inside the sensor under the external contact interactions. The contours of the contacting objects are outlined by white markers, while the contact depth characteristics can be indirectly obtained from the distribution of black pins and white markers, creating a 2.5D visualization. Based on the imaging effect, we improve the Shape from Shading (SFS) algorithm to process tactile images, enabling a coarse but fast reconstruction for the contact objects. Two experiments are performed. The first verifies TacShade's ability to reconstruct the shape of the contact objects through one image for object distinction. The second experiment shows the shape reconstruction capability of TacShade for a large panel with ridged patterns based on the location of robots and image splicing technology.

## I. INTRODUCTION

Sketches drawn with pencils can present 3D visual effects through the relationship of light, shadow and greyscale. Generally, the darker area is rendered with bold, weighty lines to generate an immersive shadow. In contrast, the lighter area emerges with fewer, feathery lines, creating a striking counter-point to the rich shadows [1]. This parallels the notion that alterations in the surface luminance of an object convey inherent 3D characteristics commonly [2]. As presented in Fig.1, a sketch drawn image uses the halftone, shadow line and form shade in a 2D plant to appear a 2.5D imaging effect to realize 3D visualization. Inspired by sketch images, in this paper, we create a new tactile sensor, called TacShade, for the creation of 2.5D tactile images that contain all the above-mentioned elements of a sketch drawing affected by the densities of points represented by markers and pins. The TacShade simulates the overlap of pencil lines by modifying the distribution of the markers and pins through interactions with the environment, emulating the changes of light and shadow effects on objects in the sketch, with the ability to indirectly acquire depth features. It constitutes a 2.5D tactile sensing method based on the Marker Displacement Method [28], which indirectly obtains depth information from a 2D image. The newly designed sensor stems from the TacTip, a single-camera tactile sensor developed by Bristol Robotics Laboratory [3], [26]. Hence, TacShade inherited advantages of TacTip e.g., low cost and high performance in pose prediction and edge following tactile servo control [4].

There are several markers that are printed on the top of papillae pins on the underside of the TacTip's soft skin. The pins amplify the deformation of the deformation of skins and cause the pins' visible markers to shift laterally. The pliable elastomer gel fills in the skin of TacTip enabling the sensor to be more resilient. Markers on the nodular pin tips are displaced laterally under the interaction with the objects and an embedded camera can observe the displacement of markers to represent and enhance the surface deformation [3]. However, the sparse nature of the markers makes TacTip challenging to use for high-resolution shape reconstruction. Different from TacTip, GelSight [5] and GelSim [6] generate shadow images of deformations from alterations through reflected light intensity. The GelSight-type sensor has high sensitivity and precision in contact position estimation and surface contour depiction of objects. However, the constraint of their 2D geometries hampers the capacity for 3D detection [7]. Despite the GelSight-type sensors endowed with their 3D shapes, such as DenseTact [7], GelSight 360 [8] and GelTip [9], their ability to estimate deep and large-area contact deformation, such as wrinkle and overlapping or depth estimation are still subject to the limitations of the thickness of the colloidal reflection layer.

To our knowledge, few existing research can deal with tactile contour rendering and 3D tactile reconstruction through in-depth deformation of tactile sensors. The TacShade was developed for this problem. Usually, the images captured from TacTip are used to estimate relative positions or positional variations of sensors with respect to the edges or surface of the object. This involves methodologies such as Bayesian probabilistic models [10], [11], support vector machines for slip detection [12], and Deep Learning for object edge tracking and tactile servo control [13]-[16]. However, the methods are abstract for 3D representation that need to build a map between the 2D images of the marker distribution and the 3D model.

The TacShade deals with 3D detection and contour recons-truction based on the new mechanical design and special image processing methods. We can read from the diagram in

* Authors contributed equally to this manuscript.
Research supported by VC-ERC project under the support of University of the West of England.

Zhenyu Lu, Yifan Li, Weiyong Si & Chenguang Yang are with the Faculty of Environment and Technology and Bristol Robotics Lab at the University of the West of England, Bristol, BS16 1QY, UK. (e-mail: Zhenyu.Lu@uwe.ac.uk, yifan2.li@live.uwe.ac.uk; Weiyong.Si@uwe.ac.uk, cyang@ieee.org).

Jialong Yang is with the School of Future Technology, South China University of Technology, Guangzhou, 511442, China and Peng Cheng Laboratory, Shenzhen, 518055, China. (e-mail: meaujlyang@mail.scut.edu.cn).

Haoran Li and Nathan F. Lepora are with the Bristol Robotics Laboratory, University of Bristol, BS8 1TW Bristol, U.K (e-mail: haoran.li@bristol.ac.uk; n.lepora@bristol.ac.uk).

† Corresponding author: Chenguang Yang

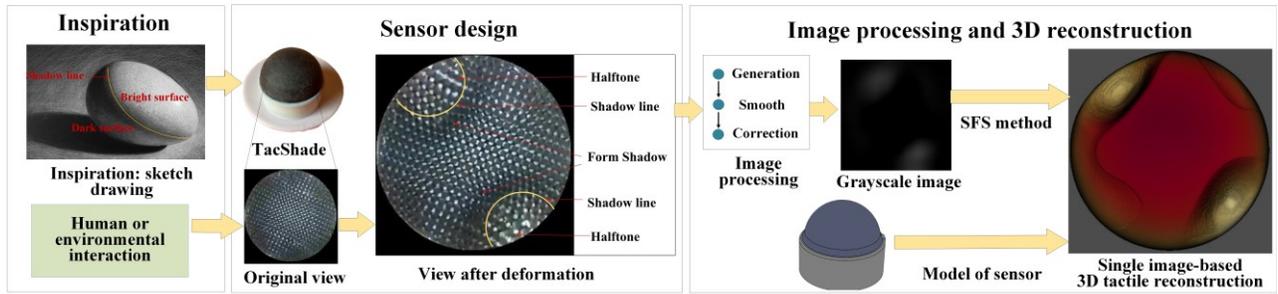

**Fig.1** Diagram of TacShade for single image-based 3D tactile reconstruction. Under the inspiration of sketch drawing, the research of TacShade consisted of two aspects: sensor design in hardware and image processing in software to achieve final tactile 3D reconstruction effect.

Fig.1 that the TacShade has no difference from the normal TacTip in appearance. Under human and outer environmental force, the skin deforms and the view-images of the inside camera change so that we can easily see the halftone, shadow lines and form shadows under the interaction of printed pins and markers. Furthermore, we develop a new convolution operation relating to the pixel area to directly transform the luminance of tactile images to greyscale images. Following this, the algorithm of SFS [19]- [22] in computer vision is then improved for a rapid tactile 3D reconstruction for both contact objects and sensor surface deformations based on a single tactile image as a new 2.5D tactile implementation. Finally, we verify the effectiveness of the new TacTip and the improved method in 3D reconstruction through two experiments.

## II. PRELIMINARY WORK OF THE TACTIP SENSOR

Since this research is developed from the TacTip Sensor, we briefly introduce the preliminary work of the TacTip for a better understanding of our design. In construction, the TacTip utilizes supple rubber materials and resilient transparent gel to fabricate bio-mimetic skin, with the purpose of emulating the epidermis and dermis of human fingertip skin. More notably, papillae pins are embedded in the rubberized epidermis, which serves as the biomimetic elements mimicking Merkel cells and intermediate ridges manifest synergistic perceptual capabilities collaboratively. TacTip employs a single camera to capture the positions of numerous pins (markers) arranged in an array-like distribution within its interior. The movements of the markers in the captured images are used to visually depict and amplify the effects of physical contact, as shown in Fig.2 (a) to Fig.2(c). The images are binarized as inputs of various learning models for enabling robots to achieve tactile perception and perform manipulation tasks.

The TacTip has developed various shapes, where the classic hemispherical shape can enhance the omnidirectional perception. The smooth, pliable surface enables it to conform to objects of various shapes and withstand greater contact depths, rendering it well-suited for 3D detection tasks [3]. TacTip has been utilized for object classification, detection and tactile servo control by using various statistics methods and Deep learning networks to achieve high accuracy for various tasks [13]-[18]. However, a dataset usually needs to collect thousands of samples, the networks should wait a long time for training and the trained models have limited applications for special duties and are hard to be transferred to other scenarios. For these challenges, the TacShade provides an intuitively presented pattern and an image processing method for 3D detection and reconstruction of the TacTip family.

## III. DESIGN CONCEPT AND FABRICATION

The TacShade maintains the advantages of the TacTip: single-camera, low cost, sensitive to dynamic actions and large surface deformation. Meanwhile, it refers to GelSight-type sensors using photometric stereo for contour and texture reconstruction, enabling it can be used for deeper 3D detection and surface reconstruction, as well as object recognition.

### A. Design Concept

In Fig. 2, we present the design concept of TacShade. Compared with the TacTip in Fig.2(a), the markers are printed by Vero White with a thickness of 0.5mm on the top of the pins. While the TacShade mosaics the markers within the gaps of the black soft pins printed by Aligus. Then, on the one hand, the density of markers increases one time and on the other hand, the functions of the markers and pins are changed. That is the markers are attached directly to the skin, then the deformation of the skin is amplified by the pins instead of being represented by the markers but by the correlations of both bins and markers.

This can be understood by the comparison between Fig.2(a) to Fig.2(c) and Fig.2(d) to Fig.2(f). For a TacTip, the markers are distributed on average. Under the pressing and squeezing interactions, the deformation on the skin is amplified to affect the distribution of the markers to outline the general effect of interactions. The TacShade utilizes the ratio of white and black for a unit region (like the green square shown in Fig.2(e) and Fig.2(f)) to reflect the interaction between objects and sensor.

Initially, the pins hide part of the white markers to achieve the preliminary white-black ratio, as Fig. 2(d) shows. Under the outer interactions, the white markers are presented more that looks brighter (Fig. 2(e)) or hidden by the pins more that looks darker (Fig. 2(f)) so that the white-black ratio in different areas will be modified to generate visual effects of halftone and form shadow. We can read the counter shape of objects from the images acquired by the TacShade (Fig.1). Compared with normal TacTip, the markers in TacShade look like providing a white background for the black pins. The deformation of skin will also be amplified by papillae pins.

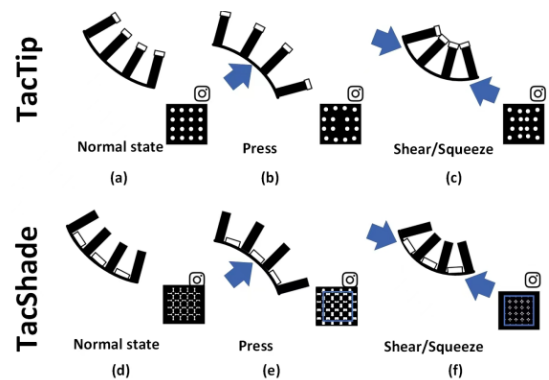

**Fig.2** Design-concept of TacShade and its compare with Tactip. (a) to (c) show the state of TacTip's pins and markers in normal state, press and shear/squeeze, and (d) to (f) show the state of the TacShade.

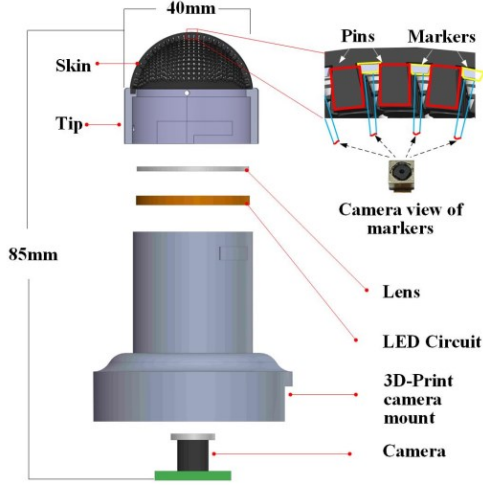

**Fig.3** Exploded diagram of TacShade, we especially show the details of the skin design that has sketching pins and hidden markers.

### B. Fabrication

Fig.3 shows the structure of the TacShade in the same fabrication as the classic TacTip. The sensor and its detection part consist of the skin and its base, named tip. The lens is used to fill up the TacTip sensor with gels and the LED Circuit is for providing light. We use a 3D-printed camera mount to fix the camera to ensure the acquired images are clear. The difference is located in the positions of the pins and the markers. In order to amplify the reflection effect, the pins and markers are specially designed in gradient cylinder-like shapes to separate the pins and markers in the 3D printing process and highlight light and shadow contrasting effects from the view of the camera, as the zoomed figure presented in Fig.3.

## IV. SHAPE RECONSTRUCTION FROM SHADING

As claimed in Section III, the TacShade reflects tactile interaction through the white-black ratio for a certain space. That inspires us to introduce the concept of greyscale to represent the white-black ratio and achieve a greyscale image. Further depth information is obtained indirectly through the greyscale. Fig. 4(a) shows a tactile image acquired by the TacShade after contact with a cube. After compressing, the concealed white markers beneath the skin of the TacShade gradually become visible. After binarization, it can be seen from Fig. 4(b) that the brightness area (a higher white-black ratio) contacting closely with the object (enclosed by a red box) is significantly outlined, as humans can observe in the eyes. From the purposes of 3D reconstruction, we can build a corresponding relationship between visual brightness and contact depth. Here, we use a specialized convolutional kernel to transform brightness variation into greyscale information to achieve a greyscale image to present deformation degradation through continuous greyscale changes. By introducing the SFS method, 3D reconstruction is realized through 3 steps in the following subsections. This approach allows rapid estimation of surface deformation degree from a single image, without the need for data acquisition to training networks.

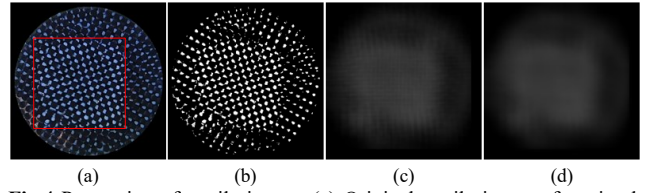

**Fig.4** Processing of tactile image: (a) Original tactile image after circular mask filtering, and the red box highlights the skin deformation area of the rectangular contact head; (b) Binary tactile image; (c) and (d) Greyscale tactile image before and after TVD filtering.

### A. Greyscale Image Generation

Within certain ranges, the larger deformation of sensor skin leads to higher brightness in the compressed area in an image. This correlation leads to an increase in the area of the white portion, denoting higher brightness. As a result, the area ratio of the white markers can be translated into a greyscale value, thus generating a greyscale image that can intuitively represent depth information. Fig. 4 presents the image processing steps. The captured original tactile image is first subjected to circular mask filtering to eliminate noise (shown in Fig.4(a)). After the greyscale conversion and binarization in Fig.4(b), we perform convolution operation on the image:

$$\hat{g}(u,v) = k^{m,n} * b(u,v) = 255 \frac{s_w^b(u,v,m,n)}{s_t^b(u,v,m,n)}, \quad (1)$$

where $(u, v)$ represents the pixel position. $b(u, v)$ and $\hat{g}(u, v)$ denote the binarized tactile image (Fig.4(b)) and the generated greyscale image after using (1) (Fig.4(c)). $k^{m,n} \in \mathbb{R}^{(m \times n)}$ represents a convolution kernel associated with the area of black and white pixels. In addition, $s_w^b$ represents the number of white pixels within a convolution window centered at $(u, v)$ and $s_t^b$ presents the amounts of the pixels within a convolution window in $b(u, v)$ respectively.

By sliding the convolutional window with a specific stride, the black and white pixel areas within the entire tactile image can be aggregated to generate a greyscale image as depicted in Fig. 4(c). To reconstruct the tactile point cloud, the crucial factor is the smoothness of depth information. However, due to the black-and-white alternating structure of the sensor, the greyscale features extracted through the sliding window are not entirely smooth. To retain the contact edge information, a Total Variation Denoising (TVD) filter is employed to smooth the greyscale and reduce noise. The smoothed greyscale image is denoted as $g(u, v)$. Compared to the Gaussian blur filter, the TVD filter exhibits superior performance in terms of both smoothing and edge preservation [23], as shown in Fig. 4(d).

### B. Contact Deformation of Skin

To transform the greyscale map into a reconstruction model, we make an analysis of the deformation of sensor skin under object compression. Fig. 5 illustrates the simplified scenario of an object-sensor interaction. The soft portion of the sensor takes a semi-spherical shape, and the coordinate system is established with the origin O at the center of the sphere. For a point S on the deformed surface, its deformation magnitude is represented by a depth along the direction toward the origin O. Here, we set the sphere's radius as r and deformation depth as $h$, and the length between S and O as $\|SO\|$. This difference is expressed as $h = r - \|SO\|$. Consequently, the coordinates of point S can be represented as $(x, y, z(x, y, h))$:

$$z(x, y, h) = \sqrt{(r-h)^2 - (x^2 + y^2)}. \quad (2)$$

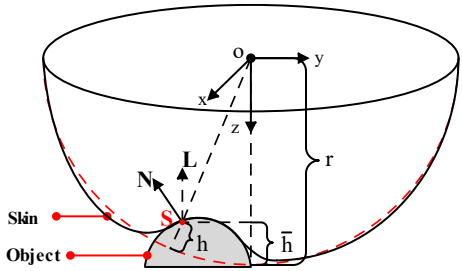

**Fig.5** Simplified contact deformation model of the TacTip (include TacShade). The red dotted line indicates the skin without contact. The degree of deformation at point S on the skin caused by contact with the object is represented by h.

Define the gradient vector $N(x, y)$ of $h(x, y)$ at S $(x, y)$ as

$$N(x, y) = (p(x, y), q(x, y), -1), \quad (3)$$

where $p = \partial h/\partial x$ and $q = \partial h/\partial y$, and the third element of $N(x, y)$ is -1, which guarantees that every point on the surface can be seen by the camera when observed from the midpoint of the camera's light in the upward direction of the z-Axis. Then if the greyscale image can depict $N(x, y)$ or deformation magnitude $h(x, y)$, the depth data can be calculated using (2).

*C. Shape Reconstruction*

In fact, the greyscale map generated in Step A has already partially conveyed the contact depth information, since it is a remarkable feature of the TacShade. A simple single-layer convolution is sufficient to roughly indicate the sensor surface deformation and magnitude. To enhance estimations of single-image greyscale depth, we introduce the SFS algorithm [20] to address the solution gradient. The SFS leverages brightness changes on the object's surface within the image to reconstruct height values, facilitating the object's shape reconstruction.

In the SFS algorithm, the object surface is usually assumed to be smooth and Lambertian, and lit by an illumination vector $\bar{L}(x, y) \in \mathbb{R}^{(1 \times 3)}$. To distinguish the symbols between the original SFS and the method used in the sensor's surface reconstruction, we use '—' for the original method and the new method doesn't have '—'. In the case of orthogonal projection, the positions $(x, y)$ in Fig.5 will be pixelated and expressed by $(u, v)$. According to the Lambertian surface reflection model under parallel illumina-tion, the following equations hold [21]:

$$\bar{g}(u,v) = \bar{R}\big(\bar{N}(u,v), \bar{L}(u,v)\big) = I\rho \frac{\bar{N}(u,v) \cdot \bar{L}(u,v)}{|\bar{N}(u,v) \cdot \bar{L}(u,v)|}, \quad (4)$$

$$\bar{N}(u,v) = (\bar{p}(u,v), \bar{q}(u,v), -1), \quad (5)$$

where $\bar{p} = \partial \bar{h}/\partial u$ and $\bar{q} = \partial \bar{h}/\partial v$, $\bar{g}(u, v)$ represents the brightness function of the Lambertian reflection surface model, $\bar{R}(*)$ is the reflection mapping function, $I$ stands for the light intensity, $\rho$ denotes the reflectance coefficient of the object's surface, and $\bar{N}(u, v)$ signifies the normal vector of the surface. If the light intensity and direction are known, $\bar{h}(u, v)$ can be calculated by (4) using SFS algorithm. And the reconstructed height $\bar{h}(u, v)$ is determined by a reference plane, such as a table surface placing the object.

However, tactile greyscale image of the TacShade is derived from a binarized image of the markers and the pins, which has two significant distinctions from the light reflection model: 1) The greyscale is affected by the camera orientation (orthogonal projection orientation, denoted as $L(u, v)$), not lighting, when illumination of the LED ring is assumed to be uniform; 2) The greyscale is primarily determined by physical mesh model of the sensor. Therefore, there are no possibilities to estimate the direction of light, and we can improve (4) by creating a new greyscale generation function $R(*)$ for the TacShade:

$$g(u, v) = R\big(N(u, v), L(u, v)\big), \quad (6)$$

where $L(u, v) = (p_c, q_c, -1) \equiv (0,0, -1)$ denotes the downward capture of markers distribution by the overhead camera of the sensor. $N(u, v)$ instead of $\bar{N}(u, v)$ in (4) shows that the hemispherical sensor shape directly affects greyscale generation. This means that the reference plane for reconstruction is the undeformed skin surface (red dotted line in Fig. 5), rather than the horizontal plane used in the original SFS algorithm. Here, we employ a linear function to approximate the function $R(*)$ in (6):

$$g(u,v) \approx \lambda \frac{N(u,v) \cdot L(u,v)}{|N(u,v) \cdot L(u,v)|} + \sigma(u,v), \quad (7)$$

where $\sigma(u, v)$ mainly stems from linear approximation errors and the errors caused by the printing precision of 3D-printer (Stratasys J826 polyjet printer), which is unmeasurable and unavoidable during production process. $\lambda$ is a constant related to the pin-markers-based greyscale generation method and the structure of the TacShade. Normalising $g(u, v)$ to obtain $g_n(u, v)$ while ignoring $\sigma(u, v)$, (7) can be transformed to

$$g_n(u,v) = R'\big(N(u,v), L(u,v)\big) \approx \frac{N(u,v) \cdot L(u,v)}{|N(u,v) \cdot L(u,v)|}, \quad (8)$$

which can be solved by the SFS algorithm. However, $g(u, v)$ is not zero even when the skin is not deformed, and there is an initial greyscale map, denoted as $g_0(u, v)$, which is caused by the volume limitation of the printed pins and markers as well as the 3D shape of the sensor. Then we calculate the greyscale variation $g_d(u, v)$ relative to $g_0(u, v)$ to determine the contact area's greyscale while reducing the impact of the greyscale in other regions:

$$g_d(u, v) = g(u, v) - g_0(u, v). \quad (9)$$

Considering that $g_0(u, v)$ represents the initial greyscale map and can reflect the shape of the TacShade to some extend. We normalize $g_d(u, v)$ to [0,1] and achieve $g_{dn}(u, v)$, as a factor to represent the modification degree of $g_0(u, v)$ under the deformation of the TacShade, then a shape-dependent greyscale $g_h(u, v)$ is achieved by

$$g_h(u, v) = g_{dn}(u, v) * g_0(u, v). \quad (10)$$

Compared to $g(u, v)$, $g_h(u, v)$ is used to express the relative deformation degree, especially for the special design of the TacShade. The processed results are shown in Figs. 6 (c) and (d). The use of $g_h(u, v)$ instead of $g_d(u, v)$ can provide a more comprehensive greyscale representation of the contact region and reduce noise. Normalizing $g_h(u, v)$ to $g_{hn}(u, v)$ and replacing $g_n(u, v)$ in (8) by $g_{hn}(u, v)$, a coarse estimation of $h$ is calculated via the hybrid linear SFS algorithm [20]. The method combines the Tsai-Shah algorithm

[21] with the Pentland algorithm [22] to avoid iterative divergence, achieving effective 3D reconstruction with fewer iterations. According to [20], $h$ can be calculated by

$$h^n(u,v) = h^{n-1}(u,v) - \frac{f(h^{n-1}(u,v))}{\frac{d}{dh(u,v)}f(h^{n-1}(u,v))}, \quad (11)$$

$$\frac{df(h^{n-1}(u,v))}{dh(u,v)} = \frac{(p+q)(pp_c + qq_c + 1)}{\sqrt{(p^2+q^2+1)^3}\sqrt{p_c^2+q_c^2+1}} - \frac{(p_c+q_c)}{\sqrt{p^2+q^2+1}\sqrt{p_c^2+q_c^2+1}}, \quad (12)$$

where $f = g_{hn} - R'$, and $n$ denotes the number of iterations, which is set as 25 in Experiments 1 and 2. $h^n(u,v)$ represents the $n$th calculation results of $h(u,v)$. The initialization and discretization methods for $h, p$ and $q$ are identical to those in [20] and will not be elaborated on here. Considering that the height estimated by the SFS algorithm is often scaled or offset from the real height of the objects, we introduce a scaling factor $\alpha$ related to the dimensions of the sensor. $\alpha$ is related to the ratio between the ground-truth contact depth and the maximum estimated depth as well as tactile image resolution, approximately set to 15 in the experiments of Section V and multiplied by $h$ to achieve the final estimated height (the 5th column of Fig. 6(e)). The reconstruction of the skin's shape can be achieved using (2) (the 6th column of Fig. 6(e)).

On a laptop with only a CPU, we implement the algorithm and process a 640×480 image in about 1.8 s. More specialised hardware and solvers can significantly increase the processing speed to achieve real-time requirements.

## V. EXPERIMENTS

This section showcases two applications of TacShade in object modeling and classification, and 3D reconstruction for large regions based on the tactile image stitching method.

### A. Experiment 1: Tactile Reconstruction for Single Objects

The experimental setup consists of a desktop robot arm, a connecting neck with several switchable contact heads and the TacShade in Fig.6(a). The robot pressure downwards on the sensor at a consistent depth but at varying positions. After capturing a tactile image (like Fig. 6(b)), a greyscale image was obtained based on (9) and (10) to get a grayscale map $g_d$ (Fig. 6(c)) and $g_h$ (Fig. 6(d)). Then we can use (11) and (2) to reconstruct the 3D shape (only the part contact with the skin) of the object and the skin of the sensor after interaction at the last two columns of Fig.6(e). Following the same way, we reconstruct the models of the other four objects and show the processed images and reconstruction models in Fig.6 (e). The outcomes show that the TacShade has different sensitivities for the objects with different sizes and shapes, even the depths and shapes of all objects are effectively depicted in the results. This indicates that the TacShade can categorize objects with just one touch, obviating the necessity to gather data with multiple contacts to train the classification network, which significantly improves classification efficiency. As the tactile reconstruction results contain noise in the non-contact region, we use the height value of the point cloud as the feature and set the number of clusters to 2. K-means clustering is then employed to extract the reconstructed object point cloud. The results are shown in the last column of Fig. 6(e). The mean error (ME) and similarity degree (SD) are calculated by

$$\text{ME} = \frac{1}{N}\sum_{\widehat{P}_i \in \widehat{P}} \min_{P_i \in P} |\widehat{P}_i - P_i|, \quad (13)$$

$$\text{SD} = 1 - \frac{d_{CD}}{h^{max}}, \quad (14)$$

where $\widehat{P}$ is the reconstructed point cloud, $P$ is the simulated ground truth and N is the number of points in $\widehat{P}$. $d_{CD}$ is the Chamfer Distance [29], which is usually used to evaluate the similarity degree of the overall shape of two point clouds. $h^{max}$ denotes the maximum contact depth of the sensor with the object. Table I shows the shape reconstruction errors. All of the MEs are within 1mm. The ball, in particular, has a very small reconstruction error. This is likely due to its curved shape, which conforms better to the soft skin of the sensor.

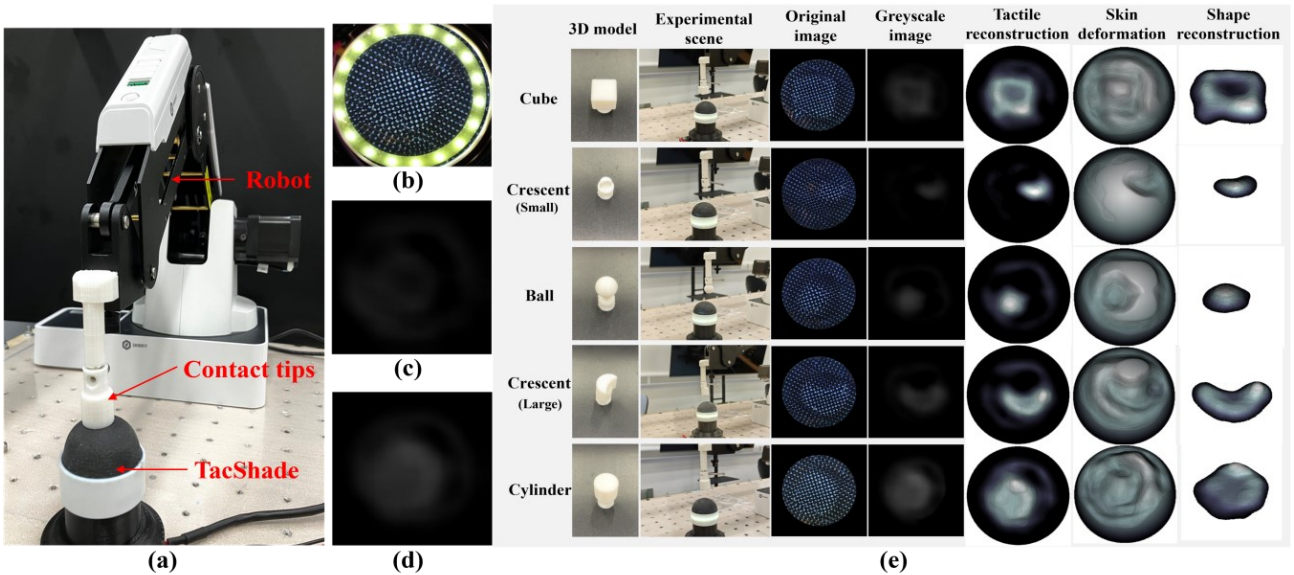

**Fig.6** Overview of the experimental setup, greyscale tactile image and 3D reconstruction results of different objects. (a) Experimental setup; (b) Tactile image; (c) Greyscale tactile image generated by $g_d$; (d) Greyscale tactile image generated by $g_h$.(e) Results of contact and reconstruction based single image. Each row represents the reconstruction results for a specific contact model, arranged from top to bottom as cube, small crescent, ball, crescent, and cylinder. Each volume shows contactscene, 3d model, original image, grayscale image, tactile reconstruction, skin deformation and shape reconstruction.

We can also find from the comparison of different objects that small objects, e.g., the small crescent at the 2nd row in Fig. 6(e), have less distinct reconstructed contours, while larger objects have clearer contours (1st, 4th and 5th rows in Fig. 6(e)). This is expected to be the reason that the precision and printing limitations of 3D printers restrict the number and size of markers, preventing high-resolution contour and texture 3D reconstruction. On the other hand, these objects with larger flat surfaces have clearer boundaries than the inner areas, resulting in a distinct section in the middle of the reconstruction model. This is a consequence of grey-scale production relying on local surface deformation (Fig. 2). When a particular contact depth is achieved, the compression of the larger level surfaces against the skin causes less discernible alterations in surface deformation in the middle region, masking the white markers beneath. Additionally, the sensor's spherical shape and vertical contact mode may result in inadequate reconstruction of the object's portion where the contact position is too close to the sensor's edge. Future research should explore ways to modify the contact direction and position to improve the sensor's omni-directional shape reconstruction capability.

TABLE I.  SHAPE RECONSTRUCTION ERRORS

| Object | Cube | Small crescent | Ball | Large crescent | Cylinder |
|---|---|---|---|---|---|
| ME (mm) | 0.9120 | 0.5008 | 0.1039 | 0.7849 | 0.8934 |
| SD (%) | 78.63 | 85.33 | 96.02 | 77.84 | 72.49 |

*B. Experiment 2: Tactile Reconstruction for A Large Region*

The TacShade is mounted at the end of a desktop robot, and the robot presses a touchpad through several contacts with the sensor, covering the entire area. We record tactile images, contact depth and corresponding robot end-effector positions for each contact. Subsequently, we perform reconstruction and get point clouds for all tactile images and stitch the point clouds using transformation matrices to get a padded model. In order to achieve a smoother point cloud, a KD tree nearest neighbour search is used to average the z-values of the points within a designated range. As the results, Fig. 7 (b) illustrates the 3D reconstruction effect and Fig. 7 (c) shows the original picture of the pad. We can obviously distinguish each featured shape with the depth information on the pad. However, although it is the first 3D reconstruction model using TacTip, improvements are still needed. For example, the results lack smoothness, and there are some discontinuities in the contour reconstruction map. Moreover, there are certain inaccuracies are observed, where the elongated shape in the upper right corner in Fig.7(b), which should have been a straight line, tilts and partially loses the point cloud. Comparable issues are also present in the case of the circular ring.

VI. DISCUSSION

Fig. 5 shows that the deformation of the sensor's skin does not perfectly conform to the surface of the object it is in contact with. Moreover, even in areas without direct contact, there are still significant deformations attributed to the soft sensor's characteristics. This imprecision leads to inaccuracy in contact model reconstruction, resulting in errors both in contact shape

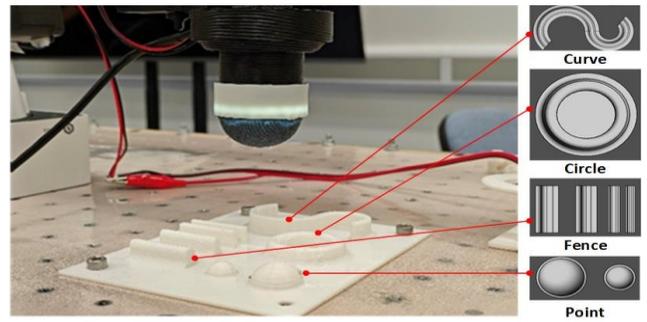

(a) Overview of experimental setup, consisting of a TacShade sensor, a Desktop robot arm (Dobot magician) and a Touchpad. This pad contains four kinds of bulges: curves, a circle, fences with different widths and intervals and two points with different radius.

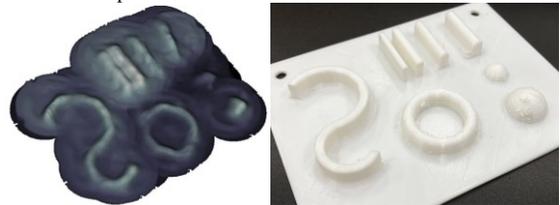

(b) Tactile 3D reconstruction result. (c) Touchpad in a photograph view.
**Fig. 7** Tactile 3D reconstruction experiment and results.

and non-contact reconstruction, as shown in the fourth column of Fig. 6(e). An object's reconstructed shape based on $h$ is an approximation. Therefore, the aim of this work is to showcase TacShade's capability to identify various objects and partially reconstruct shapes. To achieve a more precise reconstruction of the shape in contact models, it is necessary to integrate the soft sensor's deformation and shape properties. Instead of relying on the SFS based on linear approximation and artificial coefficient $\alpha$, accurate greyscale calibration and height mapping are required. Indeed, considering the use of softer materials for the sensor's soft contact head is an effective improvement. Furthermore, it's worth noting that the greyscale generated from the TacShade has a practical limitation due to the fact that the area occupied by white markers will not continue to increase beyond a certain contact depth. In future research, we will improve this design and conduct extensive investigations into this characteristic.

VII. CONCLUSION

Inspired by the principles of sketch drawings, this work introduces a newly developed tactile sensor, TacShade, which leverages the relationship between light, shadow, and grey-scale and achieves 3D shape reconstruction using a special 3D printing structure. By performing binarization and pixel-area-related convolution operations on the tactile image, and then feeding the smoothed greyscale image into the SFS algorithm, we develop a novel marker-based 2.5D tactile method to reconstruct the 3D shape of the contacted objects and describe a large pad with contour and depth information. The results illustrate the effectiveness of the new design and method in real applications. However, the results also reveal certain limitations, such as reconstruction errors and unexpected accuracy for some small objects. Future work will involve optimizing the sensors' structure, establishing an exact mapping between the greyscale and contact depth, researching tactile super-resolution abilities [24], [25], [27] and exploring the potentiality of tactile SLAM and manipulation tasks.

## REFERENCES

[1] T. C. Wang, Pencil Sketching, 2nd ed. Nashville, TN: John Wiley & Sons, 2002.

[2] C. Shao, A. Bousseau, A. Sheffer, and K. Singh, "CrossShade: Shading Concept Sketches Using Cross-Section Curves," ACM Trans. Graph., vol. 31, no. 4, Jul. 2012.

[3] B. Ward-Cherrier et al., "The TacTip Family: Soft Optical Tactile Sensors with 3D-Printed Biomimetic Morphologies," Soft Robotics, vol. 5, no. 2, pp. 216–227, 2018.

[4] N. F. Lepora, Y. Lin, B. Money-Coomes, and J. Lloyd, "DigiTac: A DIGIT-TacTip Hybrid Tactile Sensor for Comparing Low-Cost High-Resolution Robot Touch," IEEE Robotics and Automation Letters, vol. 7, no. 4, pp. 9382–9388, 2022.

[5] W. Yuan, S. Dong, and E. H. Adelson, "GelSight: High-Resolution Robot Tactile Sensors for Estimating Geometry and Force," Sensors, vol. 17, no. 12, 2017.

[6] E. Donlon, S. Dong, M. Liu, J. Li, E. Adelson, and A. Rodriguez, "GelSlim: A High-Resolution, Compact, Robust, and Calibrated Tactile-sensing Finger," in 2018 IEEE/RSJ International Conference on Intelligent Robots and Systems (IROS), 2018, pp. 1927–1934.

[7] W. K. Do and M. Kennedy, "DenseTact: Optical Tactile Sensor for Dense Shape Reconstruction," in 2022 International Conference on Robotics and Automation (ICRA), 2022, pp. 6188–6194.

[8] M. H. Tippur and E. H. Adelson, "GelSight360: An Omnidirectional Camera-Based Tactile Sensor for Dexterous Robotic Manipulation," in 2023 IEEE International Conference on Soft Robotics (RoboSoft), 2023, pp. 1–8.

[9] D. F. Gomes, Z. Lin, and S. Luo, "GelTip: A Finger-shaped Optical Tactile Sensor for Robotic Manipulation," in 2020 IEEE/RSJ International Conference on Intelligent Robots and Systems (IROS), 2020, pp. 9903–9909.

[10] N. F. Lepora, K. Aquilina, and L. Cramphorn, "Exploratory Tactile Servoing With Active Touch," IEEE Robotics and Automation Letters, vol. 2, no. 2, pp. 1156–1163, 2017.

[11] C. Yang and N. F. Lepora, "Object Exploration Using Vision and Active Touch," in 2017 IEEE/RSJ International Conference on Intelligent Robots and Systems (IROS), 2017, pp. 6363–6370.

[12] J. W. James, N. Pestell, and N. F. Lepora, "Slip Detection with a Biomimetic Tactile Sensor," IEEE Robotics and Automation Letters, vol. 3, no. 4, pp. 3340–3346, 2018.

[13] N. F. Lepora and J. Lloyd, "Optimal Deep Learning for Robot Touch: Training Accurate Pose Models of 3D Surfaces and Edges," IEEE Robotics & Automation Magazine, vol. 27, no. 2, pp. 66–77, 2020.

[14] N. F. Lepora, A. Church, C. de Kerckhove, R. Hadsell, and J. Lloyd, "From Pixels to Percepts: Highly Robust Edge Perception and Contour Following Using Deep Learning and an Optical Biomimetic Tactile Sensor," IEEE Robotics and Automation Letters, vol. 4, no. 2, pp. 2101–2107, 2019.

[15] N. F. Lepora and J. Lloyd, "Pose-Based Tactile Servoing: Controlled Soft Touch Using Deep Learning," IEEE Robotics & Automation Magazine, vol. 28, no. 4, pp. 43–55, 2021.

[16] W. Fan, M. Yang, Y. Xing, N. F. Lepora, and D. Zhang, "Tac-VGNN: A Voronoi Graph Neural Network for Pose-Based Tactile Servoing," in 2023 IEEE International Conference on Robotics and Automation (ICRA), 2023, pp. 10373–10379.

[17] Y. Lin et al., "Bi-Touch: Bimanual Tactile Manipulation with Sim-to-Real Deep Reinforcement Learning," IEEE Robotics and Automation Letters, vol. 8, no. 9, pp. 5472–5479, 2023.

[18] J. Lloyd and N. F. Lepora, "Goal-Driven Robotic Pushing Using Tactile and Proprioceptive Feedback," IEEE Transactions on Robotics, vol. 38, no. 2, pp. 1201–1212, 2022.

[19] Z. Cao et al., "The Algorithm of Stereo Vision and Shape from Shading Based on Endoscope Imaging," Biomedical Signal Processing and Control, vol. 76, pp. 103658, 2022.

[20] M. Kotan, C. Öz, and A. Kahraman, "A Linearization-based Hybrid Approach for 3D Reconstruction of Objects in A Single Image," International Journal of Applied Mathematics and Computer Science, vol. 31, no. 3, pp. 501–513, 2021.

[21] T. Ping-Sing and M. Shah, "Shape from Shading Using Linear Approximation," Image and Vision Computing, vol. 12, no. 8, pp. 487–498, 1994.

[22] A. Pentland, "Shape Information from Shading: A Theory About Human Perception," in [1988 Proceedings] Second International Conference on Computer Vision, 1988, pp. 404–413.

[23] M. Nikolova, "An Algorithm for Total Variation Minimization and Applications," J. Math. Imaging Vis., vol. 20, no. 1/2, pp. 89–97, 2004.

[24] H. Sun and G. Martius, "Guiding the design of superresolution tactile skins with taxel value isolines theory," Science Robotics, vol. 7, no. 63, p. eabm0608, 2022.

[25] Y. Yan et al., "Soft Magnetic Skin for Super-resolution Tactile Sensing with Force Self-decoupling," Science Robotics, vol. 6, no. 51, p. eabc8801, 2021.

[26] N. F. Lepora, "Soft Biomimetic Optical Tactile Sensing with the TacTip: A Review," IEEE Sensors Journal, vol. 21, no. 19, pp. 21131–21143, 2021.

[27] N. F. Lepora, U. Martinez-Hernandez, M. Evans, L. Natale, G. Metta, and T. J. Prescott, "Tactile Superresolution and Biomimetic Hyperacuity," IEEE Transactions on Robotics, vol. 31, no. 3, pp. 605–618, 2015.

[28] M. Li, T. Li, and Y. Jiang, "Marker Displacement Method Used in Vision-Based Tactile Sensors—From 2-D to 3-D: A Review," IEEE Sensors Journal, vol. 23, no. 8, pp. 8042–8059, 2023.

[29] H. Fan, H. Su, and L. Guibas, "A point set generation network for 3D object reconstruction from a single image," in 2017 IEEE Conference on Computer Vision and Pattern Recognition (CVPR), 2017, pp. 2463-2471.